\documentclass[final,5p,times,twocolumn]{elsarticle}

\usepackage{amssymb,amsmath}
\usepackage{textcomp}
\usepackage{subfigure}
\usepackage{braket}
\usepackage{float}
\usepackage{xcolor}
\usepackage{ulem}
\usepackage{cleveref}
\usepackage{multirow}
\usepackage{icomma}
\usepackage{stackrel}
\usepackage{appendix}
\usepackage{booktabs}

\begin{document}

\title{Comparative study of the heavy-quark dynamics with the Fokker-Planck Equation and the Plastino-Plastino Equation}

\author{Eugenio Meg\'{\i}as}
 \address{Departamento de Física Atómica, Molecular y Nuclear and Instituto Carlos I de Física Teórica y Computacional, \\
 Universidad de Granada, Avenida de Fuente Nueva s/n, 18071 Granada, Spain
 }
\ead{emegias@ugr.es}

\author{Airton Deppman}
 \address{Instituto de Física -  Universidade de São Paulo, Rua do Matão 1371, São Paulo 05508-090, Brazil
 }
\ead{deppman@usp.br}

\author{Roman~Pasechnik}
\address{Department of Physics, Lund University, S\"olvegatan 14A,
Lund SE-22362, Sweden}
\ead{Roman.Pasechnik@hep.lu.se} 

\author{Constantino Tsallis}
 \address{Centro Brasileiro de Pesquisas Fisicas and National Institute of Science and Technology of Complex Systems, \\Rua Xavier Sigaud 150, Rio de Janeiro-RJ 22290-180, Brazil \\ Santa Fe Institute, 1399 Hyde Park Road, Santa Fe, 87501 NM, USA\\
 Complexity Science Hub Vienna - Josefst\"adter Strasse 39, 1080 Vienna, Austria}
\ead{tsallis@cbpf.br}

\begin{abstract}
 The Fokker-Planck Equation (FPE) is a fundamental tool for the investigation of kinematic aspects of a wide range of systems. For systems governed by the non-additive entropy $S_q$, the Plastino-Plastino Equation (PPE) is the correct generalization describing the kinematic evolution of such complex systems. Both equations have been applied for investigations in many fields, and in particular for the study of heavy quark evolution in the quark-gluon plasma. In the present work, we use this particular problem to compare the results obtained with the FPE and the PPE, and discuss the different aspects of the dynamical evolution of the system according to the solutions for each equation. The comparison is done in two steps, first considering the modification that results from the use of a different partial derivative equation with the same transport coefficients, and then investigating the modifications by using the non-additive transport coefficients. We observe clear differences in the solutions for all the cases studied here and discuss possible experimental investigations that can indicate which of those equations better describes the heavy-quark kinematics in the medium. The results obtained here have implications in the study of anomalous diffusion in porous and granular media, in Cosmology and Astrophysics. The obtained results reinforce the validity of the relation $(q-1)^{-1}=(11/3)N_c-(4/3)(N_f/2)$, where $N_c$ and $N_f$ are, respectively, the number of colours and the effective number of flavours. This equation was recently established in the context of a fractal approach to QCD in the non-perturbative regime.
\end{abstract}


\maketitle


The interest in heavy-quark dynamics as a tool to probe the quark-gluon plasma has increased with the High Energy Physics (HEP) experiments showing that the interaction of those particles with the medium is relevant, as evidenced by the observation of jet-quenching and of the collective flow observed through hadrons with charm or heavier quarks. Those observations have placed the Fokker-Planck Equation (FPE) at the centre of the theoretical investigations on the heavy-quark dynamics in the medium~\cite{Hees-Rapp}. This equation is a second-order approximation of the Boltzmann Equation, but it is preferred since it avoids the integral equation by the introduction of two transport coefficients, the drift term and the diffusion term. A review of the applications of the FPE in HEP can be found in Ref.~\cite{He:2022ywp}.

The procedure to obtain the FPE from the BE is well known~\cite{Das-Alam-Mohanty}, and it leads to the second-order differential equation
\begin{equation}
 \frac{\partial f}{\partial t}- \frac{\partial}{\partial p_i} \left[A_i(\mathbf p) f+ \frac{\partial }{\partial p_j} \left( B_{ij}(\mathbf p) f\right) \right] = 0 \,. \label{FPE}
\end{equation}
Here and below, we use natural units $\hbar = c =k_B= 1$, unless noted otherwise. The transport coefficients $A_i$ and $B_{ij}$ have been obtained from microscopic calculations~\cite{Svetitsky:1987gq}, establishing a clear path for studying the heavy-quark dynamics. The solution $f \equiv f(\mathbf p,t)$ is the momentum distribution of the heavy quark.

The bulk matter formed at high-energy collisions presents evidence that 
nonextensive statistical mechanics (based on the nonadditive entropy S$_q$)~\cite{Tsallis} is the correct framework for studying the thermostatistical properties of the quark-gluon plasma (QGP). Many analyses of the experimental data have been produced in the use of the S$_q$ with success to describe multiparticle production that results from the decay of the QGP~\cite{Marques-Cleymans-Deppman-2015,Marques-Andrade-Deppman-2013}. The clearest evidence of S$_q$ is the heavy-tailed momentum distributions observed, which can be accurately described by a $q$-exponential curve in a wide range of transverse momentum and several decades of cross-section, showing the non-extensive behaviour predicted by S$_q$~\cite{WilkWlodarkzyk-multiparticle}. This different statistical framework imposes the use of the non-extensive thermodynamics~\cite{TsallisBook} to describe the thermal properties of the medium.

The non-extensive thermodynamics of the QGP has its impacts also in the heavy-quark dynamics. In systems governed by S$_q$, the correct dynamical equation is not the FPE, but the Plastino-Plastino Equation (PPE)~\cite{PLASTINO1995347}, which has the form 
\begin{equation}
 \frac{\partial f}{\partial t}- \frac{\partial}{\partial p_i} \left[A_i(\mathbf p) f+ \frac{\partial }{\partial p_j} \left( B_{ij}(\mathbf p) f^{2-q} \right) \right] = 0 \,. \label{PPE}
\end{equation}
This equation is a generalization of the FPE that reduces to the standard FPE when the entropic index $q=1$, otherwise it differs from the standard equation. In the present work, we compare the dynamics of the heavy quark in the medium as obtained by the FPE and by the PPE. For this comparison, we compute the transport coefficients based on microscopic calculations and use the same coefficients to find the solutions of the FPE and PPE, comparing the results in both cases.

Eqs.~(\ref{FPE}) and~(\ref{PPE}) correspond, in momentum space, to the Fokker-Planck and Plastino-Plastino equations in the coordinate space. Observe that, in the absence of external forces, the drift coefficient is $A_i=0$, and Eq.~(\ref{FPE}) reduces to the heat equation, first proposed by Joseph Fourier in 1822. Under the same conditions, Eq.~(\ref{PPE}) reduces to the porous media diffusion equation~\cite{Muskat}, which has been investigated in Refs.~\cite{Schwammle,Schwammle2009}.



For the microscopic calculation of the transport coefficients we follow Ref.~\cite{Svetitsky:1987gq}, and define double-average of an undetermined function $F(p')$ as
\begin{equation}
 \begin{split}
    \langle\langle F(p') \rangle \rangle= & \frac{1}{2\omega_p}  \int \frac{d^3p'}{(2\pi)^3 2\omega_{p'}} \int \frac{d^3q}{(2\pi)^3 2\omega_{q}} \int \frac{d^3q'}{(2\pi)^3 2\omega_{q'}} |{\cal M} |^2 \times \\ & \delta^4\left((\mathbf p'-\mathbf p)-(\mathbf q-\mathbf q') \right) f(q) F(p')\,, \label{doubleaverage}
 \end{split}
\end{equation}
where the matrix elements $|{\cal M}|^2$ are calculated in second-order approximation. The coefficients can be calculated from the expression above by
\begin{equation}
 \begin{cases}
  A_i(\mathbf p) = \langle \langle (p-p')_i \rangle \rangle \\
  \\
  B_{ij}(\mathbf p)  = \frac{1}{2} \langle \langle (p-p')_i (p-p')_j\rangle \rangle 
 \end{cases} \,.
\end{equation}

The calculation of the transport coefficients can be facilitated by defining $A(p^2)$, $B_0(p^2)$ and $B_1(p^2)$ such that
\begin{equation}
 A(p^2) = \frac{p_iA_i(\mathbf p)}{p^2} =\langle \langle 1 \rangle \rangle - \frac{\langle \langle \mathbf p \cdot \mathbf p' \rangle \rangle}{p^2} \,,
\end{equation}
and
\begin{equation}
 B_0(p^2) =\frac{1}{2}\left[\delta_{ij}-\frac{p_ip_j}{p^2}\right] B_{ij}(\mathbf p) \,, \quad B_1(p^2) =\frac{p_ip_j}{p^2}B_{ij}(\mathbf p)  \,,
\end{equation}
where $p^2 \equiv |\mathbf p|^2$. Then, the coefficients may be decomposed as
\begin{eqnarray}
A_i({\mathbf p}) &=& p_i \, A(p^2) \,, \label{eqA}\\
B_{ij}({\mathbf p}) &=& \left( \delta_{ij} - \frac{p_i p_j}{p^2} \right) B_0(p^2) + \frac{p_i p_j}{p^2} B_1(p^2) \,. \label{eqB}
\end{eqnarray}
For the computation of the coefficients using the matrix elements given in Ref.~\cite{Svetitsky:1987gq}, we need to make a Lorentz transformation to the centre-of-momentum (CM) frame. This is done by using the transformation
\begin{equation}
 \begin{cases}
  \mathbf{\hat p}=\gamma_{cm}(\mathbf p - \mathbf v_{cm}E) \\
  \\
  \hat E=\gamma_{cm}(E-\mathbf v_{cm}\cdot \mathbf p)
 \end{cases}\,, \label{eq:Lorentz}
\end{equation}
where $\gamma_{cm} \equiv 1/\sqrt{1-\mathbf v_{cm}^2}$. The first line in Eq.~(\ref{eq:Lorentz}) is valid in the approximation that $\mathbf v_{cm}$ is parallel to $\mathbf p$.

The description above shows the main ingredients in the computation of the transport coefficients. Notice that we have multiplied the gluons matrix elements by a factor $-1/(4\pi)$, and those for quarks by $1/(2\pi)$. These modifications take into account the information, in Ref.~\cite{Svetitsky:1987gq}, of a multiplicative factor in the calculation of the matrix elements.


The coefficients $B_0(p^2)$ and $B_1(p^2)$ obtained in the last sections differ by a negligible amount, so we use the approximation $B_0(p^2)=B_1(p^2) \equiv B(p^2)$, and in this case we have $B_{ij}({\mathbf p}) =\delta_{ij}B(p^2)$. We also notice that $A(p^2)$ and $B(p^2)$ are slowly increasing with $p$ in the relevant region of the present study, so we approximate both to constants $A$ and $B$, respectively. These approximations were discussed in some detail in Ref.~\cite{WaltonRafelski}.

For the 3-dimensional solution of the FPE we use the ansatz
\begin{equation}
 f(\mathbf p,t)=\frac{1}{\left(\sqrt{2\pi} \sigma(t)\right)^3} ~~\exp\left[-\frac{(\mathbf p - \mathbf p_M(t))^2}{2 \sigma(t)^2} \right] \,.  \label{eq:f}
\end{equation}
This general function is a solution of the FPE if
\begin{equation}
 \begin{cases}
  \mathbf p_{M}(t)=\mathbf p_{o} \exp[-A t] \\
  \\
\sigma(t) = \sigma_o \sqrt{ \left( 1 - \frac{B}{A\sigma_0^2}\right) \exp[-2At] + \frac{B}{A\sigma_0^2}} 
 \end{cases}\,, \label{soltutionparameters}
\end{equation}
where $\mathbf p_{o}$ is a constant vector corresponding to the heavy-quark initial momentum, while $\sigma_o$ is the initial width of the distribution\footnote{The second line of Eq.~(\ref{soltutionparameters}) can be written also in the form
\begin{equation}
 \sigma(t)= \sqrt{\frac{B}{A} \left( 1 - \exp[-2A(t+\lambda)] \right) } \,,
\end{equation}
where the constant $\lambda$ can be determined by the initial condition $\sigma_o \equiv \sigma(t=0)= \sqrt{\frac{B}{A}\left( 1 - \exp[-2A\lambda] \right)}$.}. The width in equilibrium is $\sigma_\infty \equiv \lim_{t\to\infty} \sigma(t) = \sqrt{B/A}$. Notice that $\sigma_o$ can have any positive value, even larger than $\sigma_\infty$.

\begin{figure*}[t]
\centering
\includegraphics[width=0.30\textwidth]{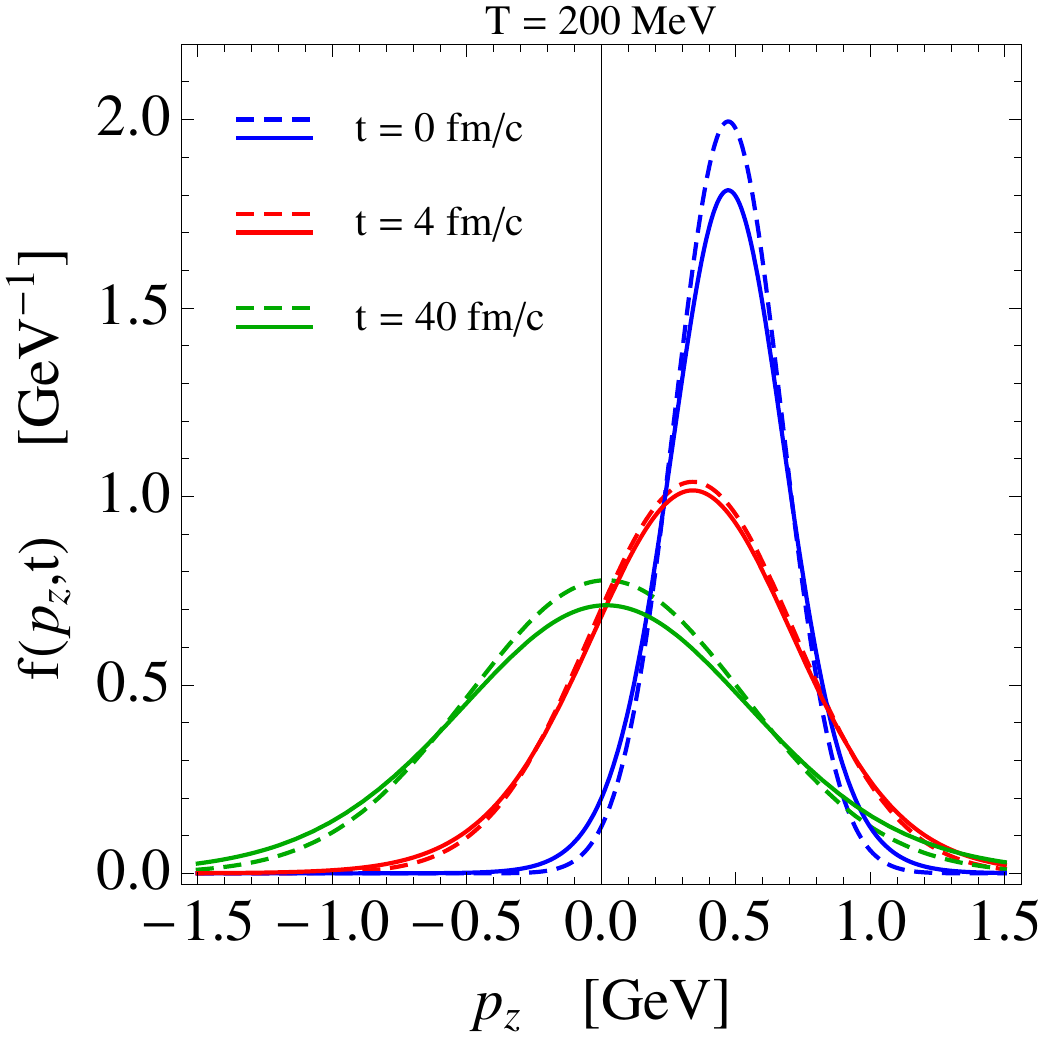}  \hspace{0.5cm} \includegraphics[width=0.30\textwidth]{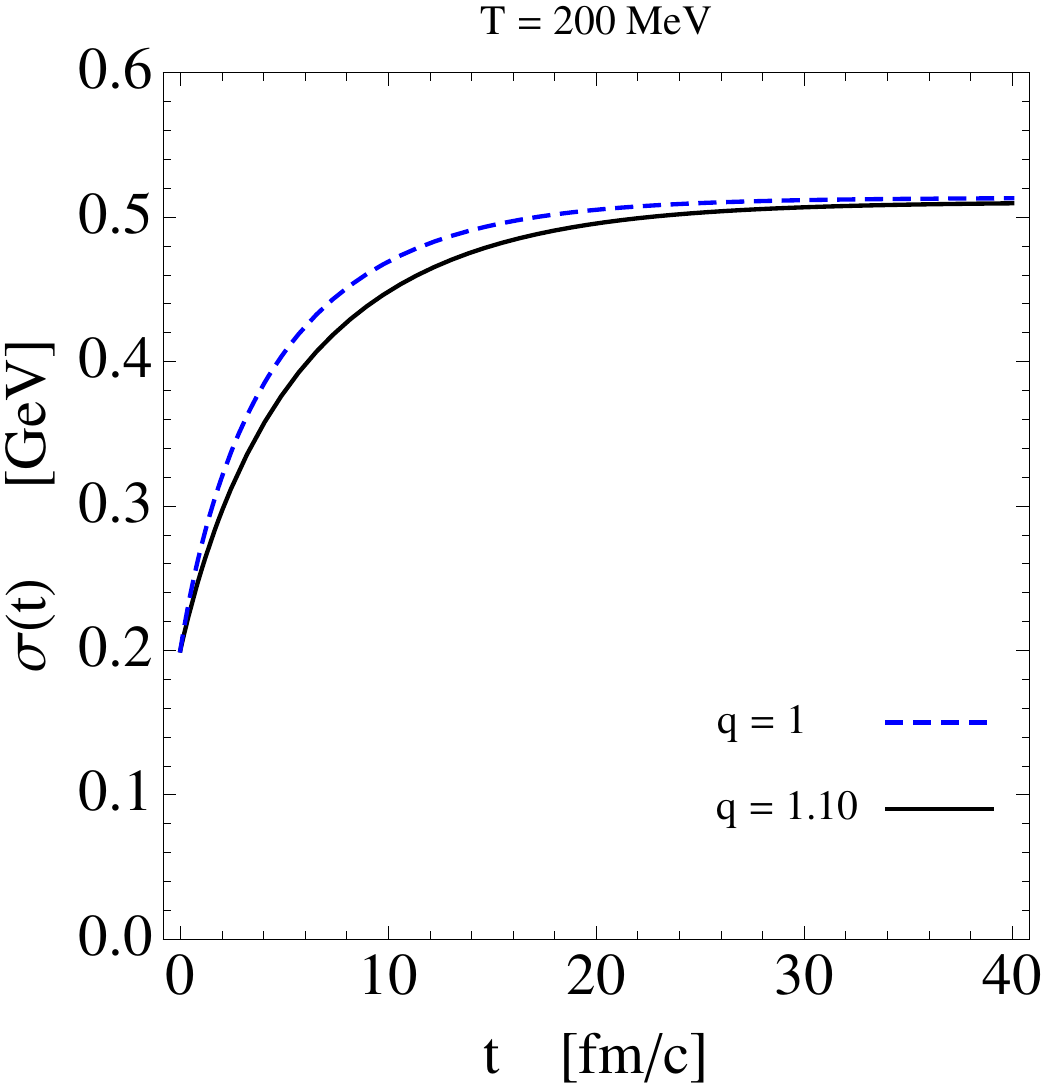}  \hspace{0.5cm} \includegraphics[width=0.30\textwidth]{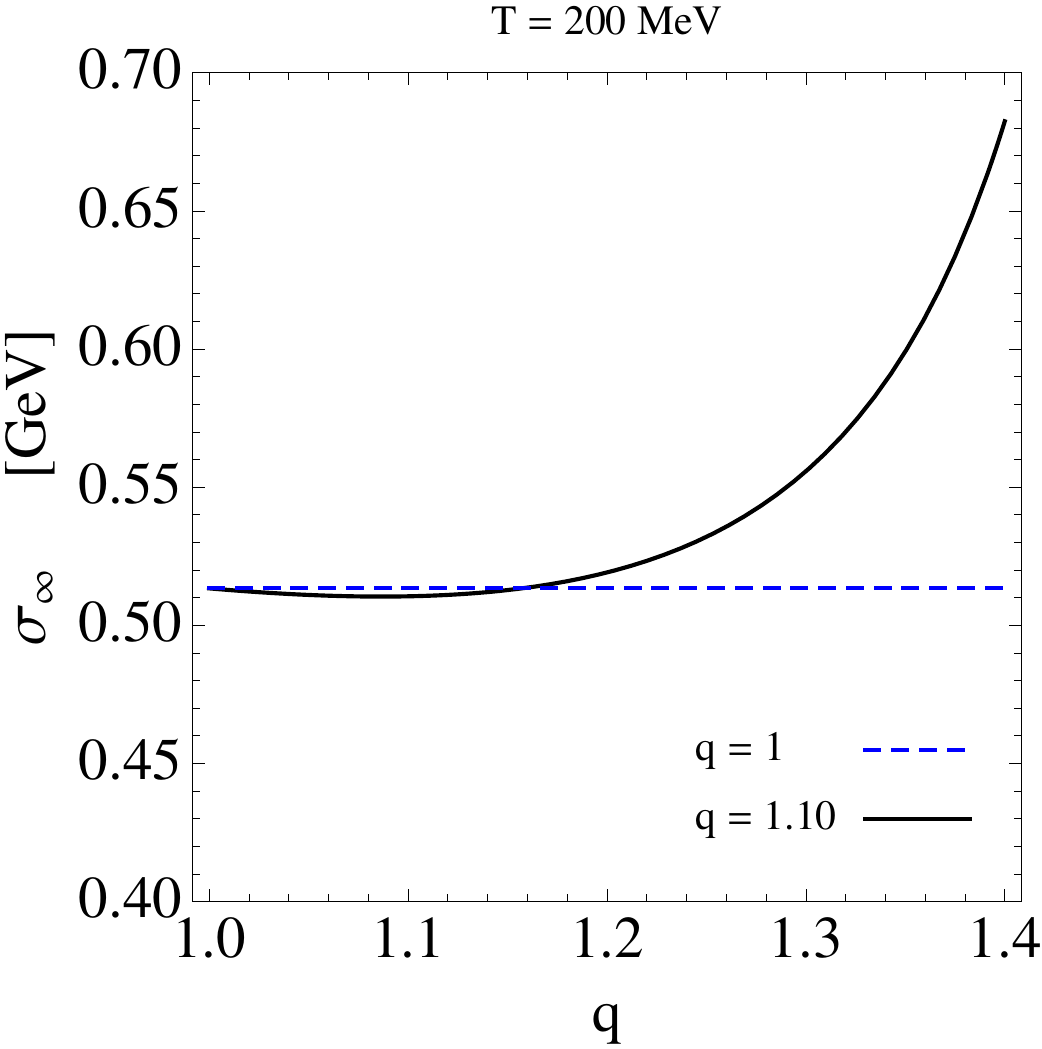}
\includegraphics[width=0.30\textwidth]{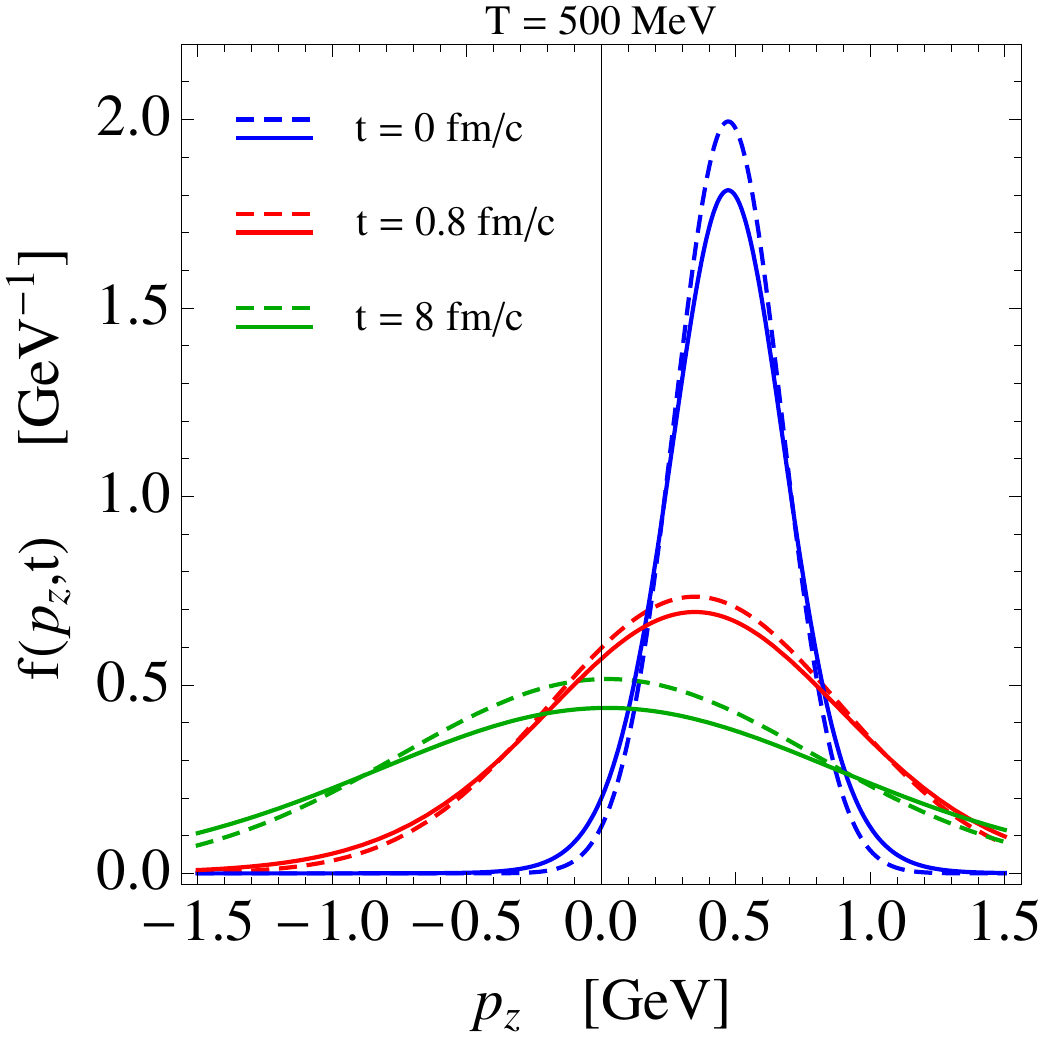}  \hspace{0.5cm} \includegraphics[width=0.30\textwidth]{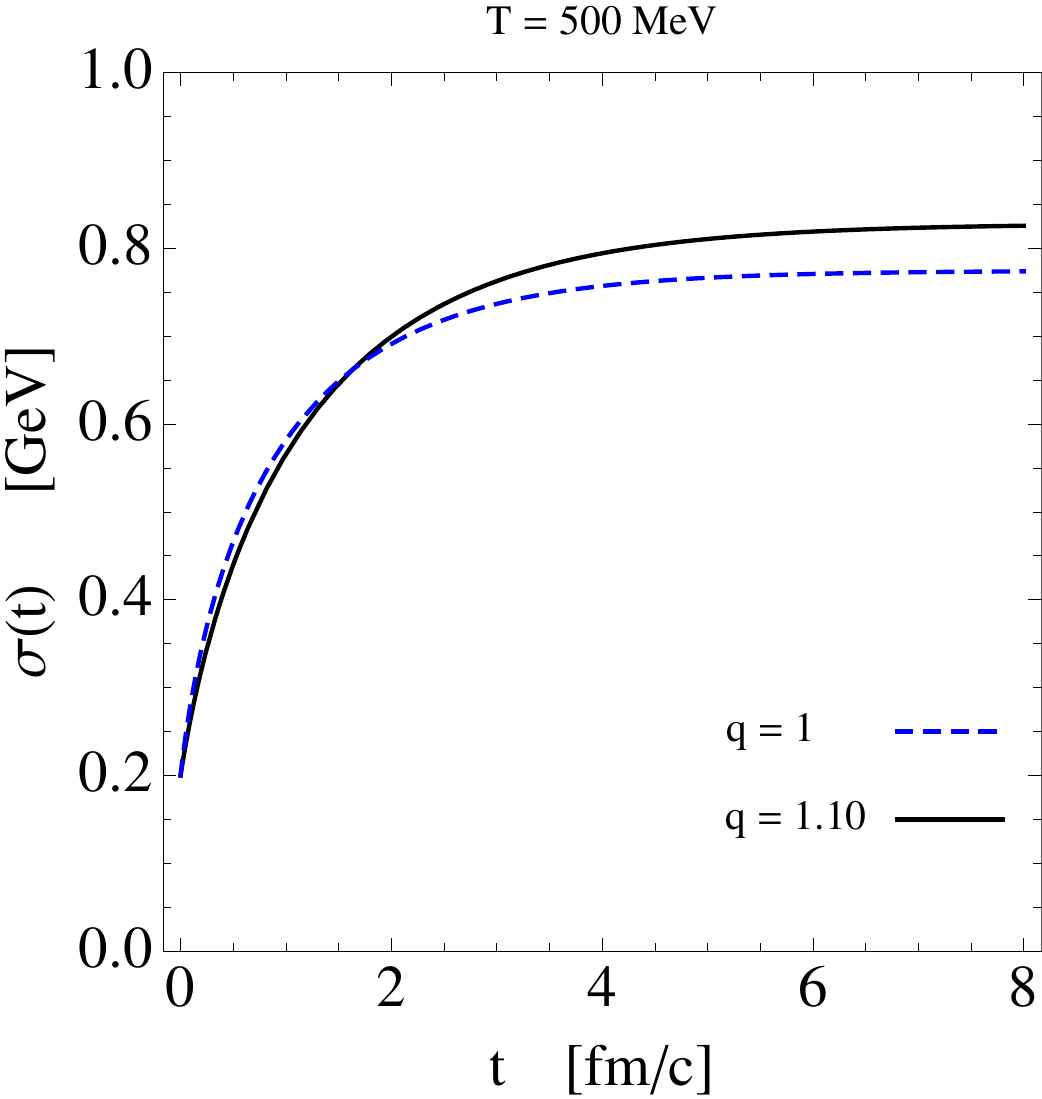}  \hspace{0.5cm} \includegraphics[width=0.30\textwidth]{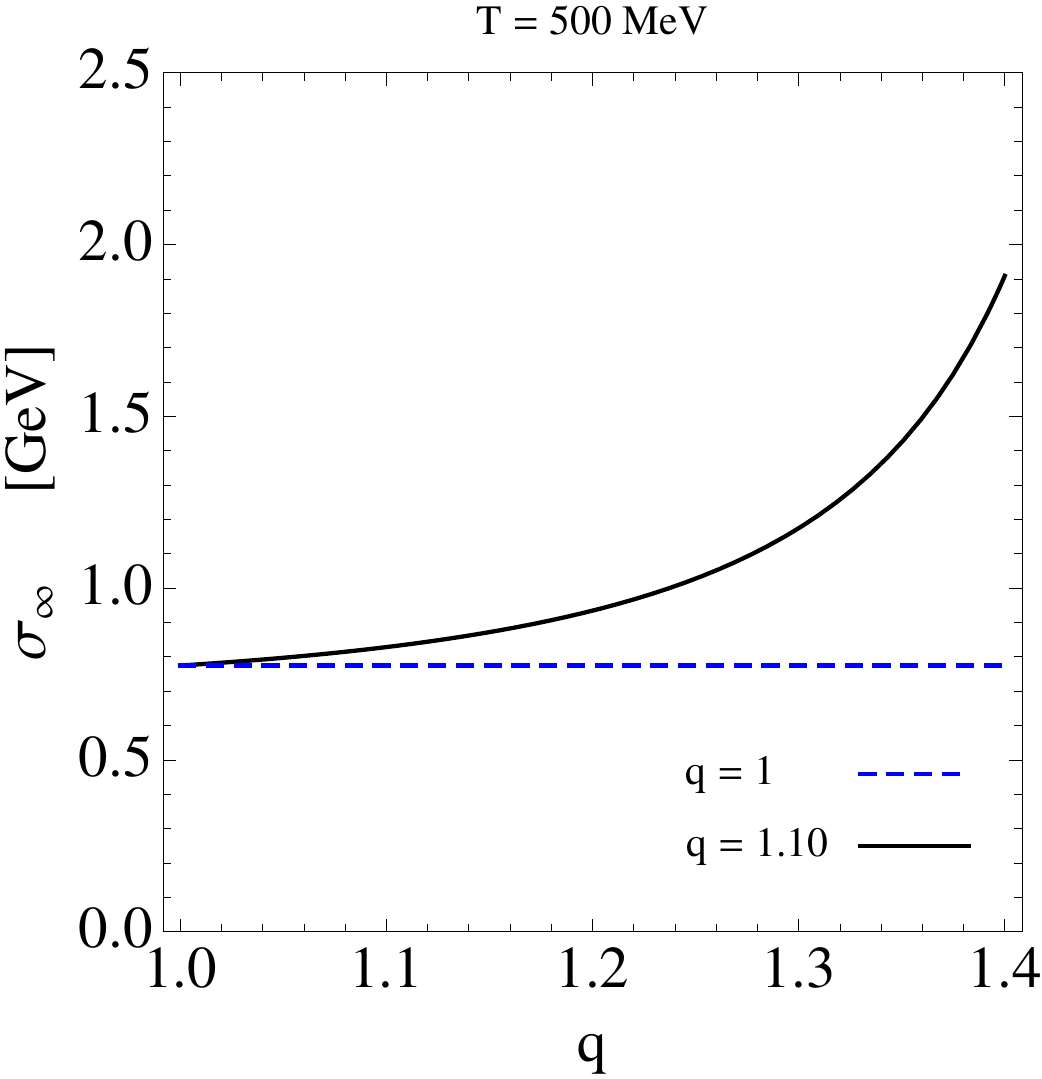}
 \caption{Left panels: Plots of the distribution functions integrated out in transverse momenta, $f(p_z,t) \equiv \int_{-\infty}^\infty dp_x dp_y f(\mathbf p,t)$, as a function of $p_z$. We display the distributions for several values of $t$. Middle panels: Widths of the distributions as a function of time. Right panels: Widths of the distributions in equilibrium, $\sigma_\infty \equiv \lim_{t\to\infty}\sigma(t)$, as a function of $q$. In these panels, solid lines correspond to $q = 1.10$ and dashed lines to $q = 1$. We have considered the initial momentum for the heavy quark to be $p_o = 0.47\, \textrm{GeV}$ in the $z$-direction, the initial width $\sigma_o = 0.2 \, \textrm{GeV}$, while $(A = 0.0820 \,\textrm{fm}^{-1}, B = 0.0216 \,\textrm{GeV}^2/\textrm{fm})$ for $T=200 \,\textrm{MeV}$ (upper panels) and  $(A = 0.381 \,\textrm{fm}^{-1}, B = 0.228 \,\textrm{GeV}^2/\textrm{fm})$ for $T=500 \,\textrm{MeV}$ (lower panels).}
\label{fig:fpz}
\end{figure*}

For the 3-dimensional solution of the PPE we use the ansatz~\footnote{The $q$-exponential function is defined as $\exp_q[x] = \left[ 1 - (q-1) x\right]^{-\frac{1}{q-1}}$.}
\begin{equation}
 f(\mathbf p,t)=  \frac{1}{\left(\sqrt{2\pi\chi_q}  \sigma(t) \right)^3} ~~\exp_q\left[-\frac{(\mathbf p - \mathbf p_M(t))^2}{2 \sigma(t)^2} \right] \,, \label{eq:fq}
\end{equation}
which is a solution of the PPE if the parameter $\mathbf p_M(t)$ is given by the first line of Eq.~(\ref{soltutionparameters}), while $\sigma(t)$ is
\begin{equation}
\sigma(t) = \sigma_o\left[ \left(1 - \kappa  \right) \exp\left[-(5-3q)At \right]  + \kappa  \right]^{\frac{1}{5-3q}} \,,
\end{equation}
with
\begin{equation}
\kappa \equiv (2-q) \left( 2\pi \chi_q\right)^{\frac{3}{2}(q-1)} \frac{B}{A} {\sigma_o^{3q-5} } \,,
\end{equation}
and
\begin{equation}
\chi_q \equiv \frac{1}{q-1} \left( \frac{\Gamma\left( \frac{1}{q-1} - \frac{3}{2}\right) }{\Gamma\left( \frac{1}{q-1} \right)} \right)^{\frac{2}{3}} \,.
\end{equation}
The width of the distribution in equilibrium
\begin{equation}
\sigma_\infty = \sigma_0 \kappa^{\frac{1}{5-3q}} = \left[ (2-q) \left( 2\pi \chi_q\right)^{\frac{3}{2}(q-1)} \frac{B}{A} \right]^{\frac{1}{5-3q}} \label{sigmainft}
\end{equation}
is independent of $\sigma_o$, and it diverges for $q\to 5/3$. It is straightforward to see that the solution of the PPE given by Eq.~(\ref{eq:fq}) tends to the solution of the FPE of Eq.~(\ref{eq:f}) in the limit $q \to 1$, and in particular $\chi_q \stackbin[q \to 1]{}{\to} 1$. Both distributions are normalized as 
\begin{equation}
\int d^3p f(\mathbf p,t) = 1 \,.
\end{equation}
We display in Fig.~\ref{fig:fpz} (left) the distributions obtained with the FPE and the PPE for two different values of the medium temperature: $T=200\,\textrm{MeV}$ (upper panels) and $500\,\textrm{MeV}$ (lower panels). The behaviours of the widths 
as functions of time are displayed in the middle and right panels of this figure for both $q=1.10$ (S$_q$) and $q=1$ (Boltzmann). The values of the constants $A$ and $B$ have been computed, and they turn to be in agreement with those shown in Ref.~\cite{Svetitsky:1987gq}.

\begin{figure*}[t]
\centering
\includegraphics[width=0.30\textwidth]{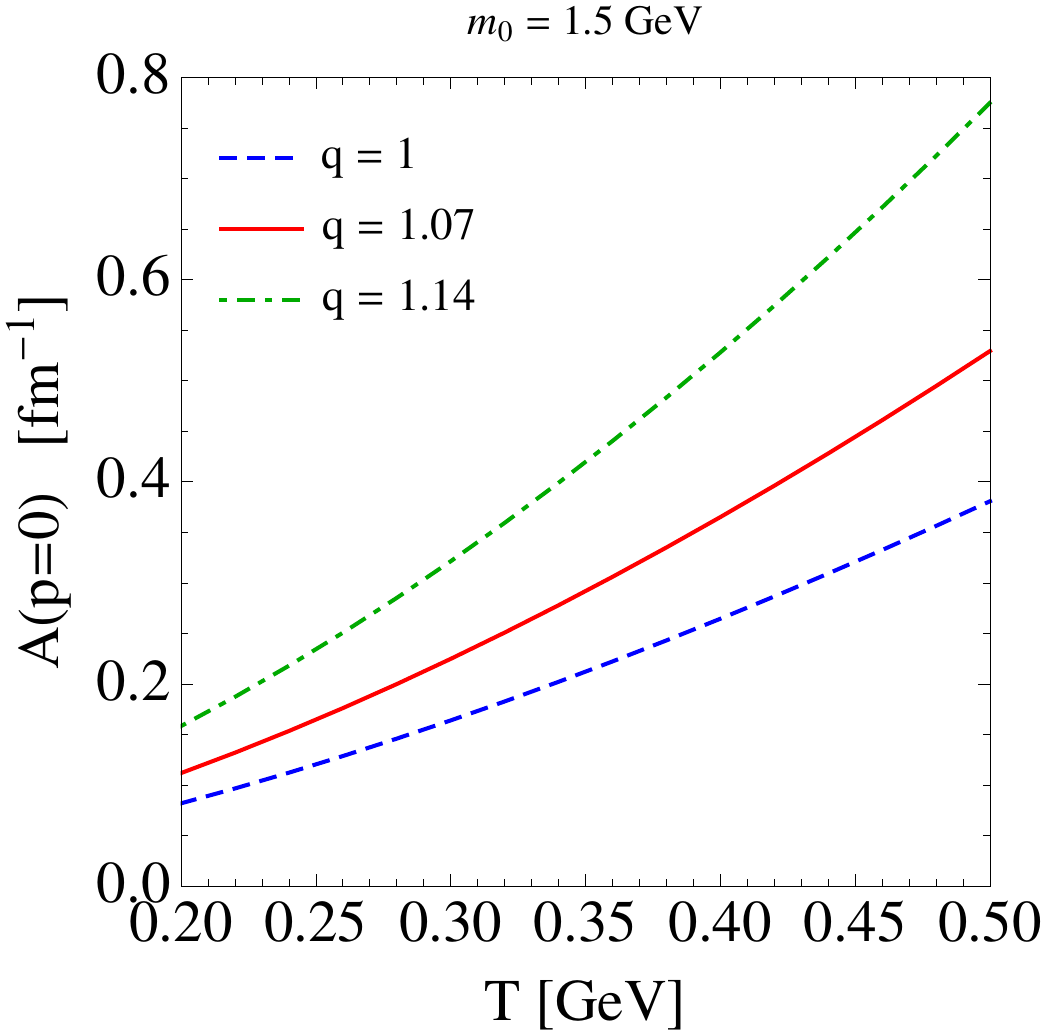}  \hspace{0.5cm} 
\includegraphics[width=0.31\textwidth]{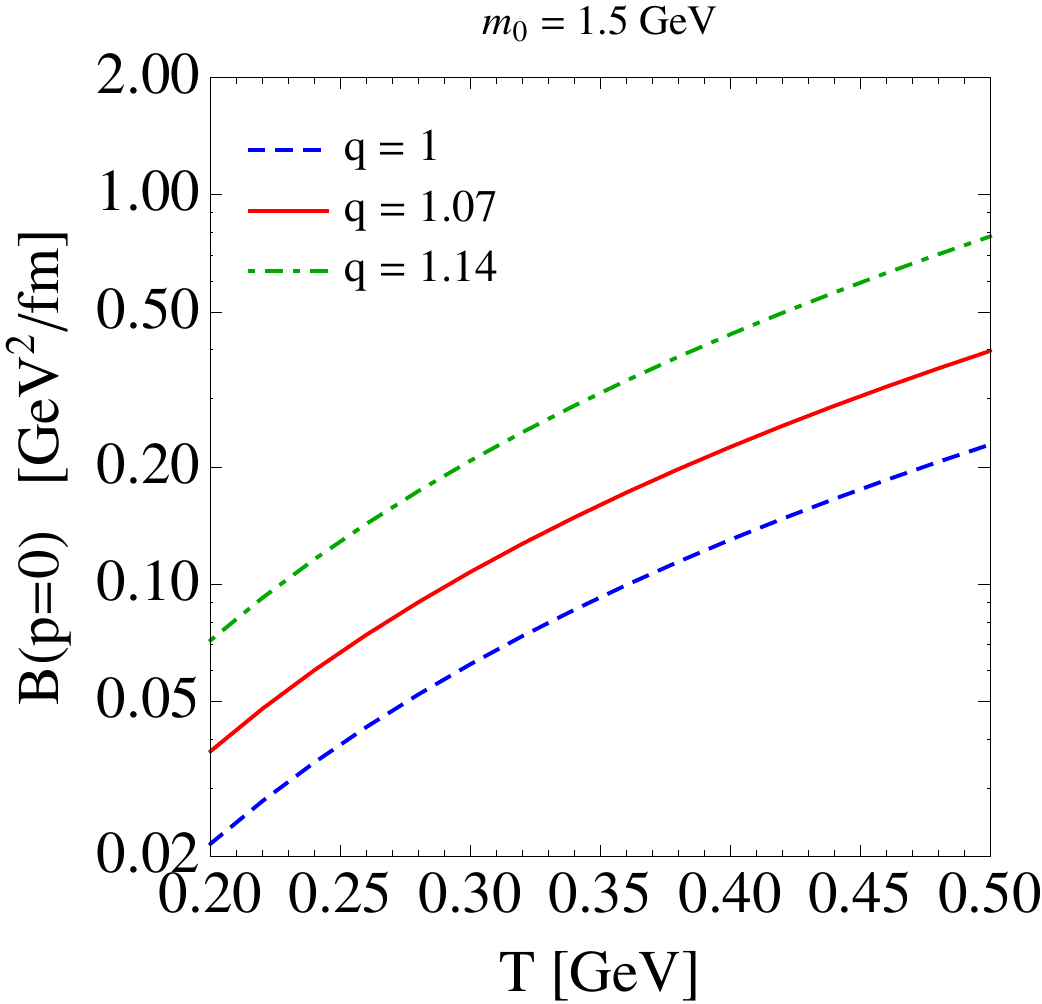}  \hspace{0.5cm} \includegraphics[width=0.30\textwidth]{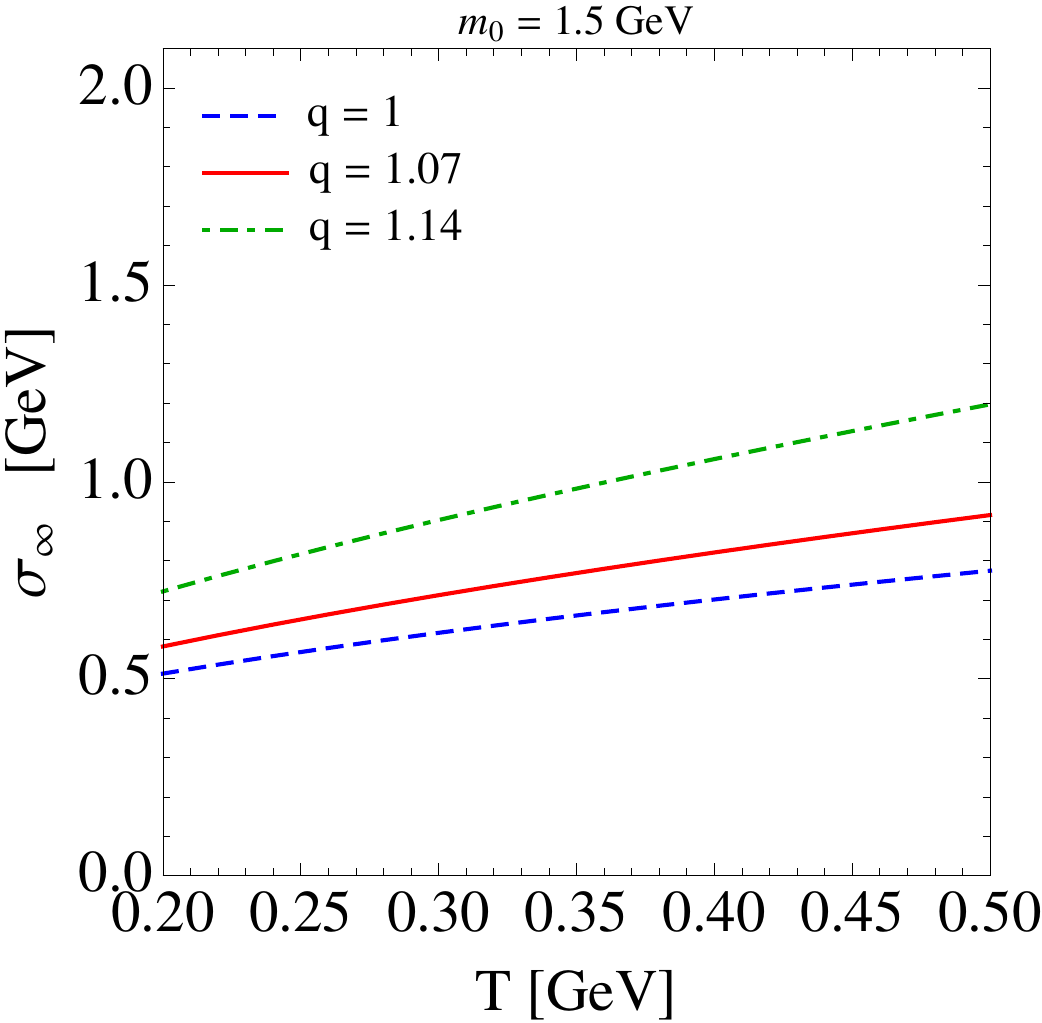} \hspace{0.5cm}
\includegraphics[width=0.30\textwidth]{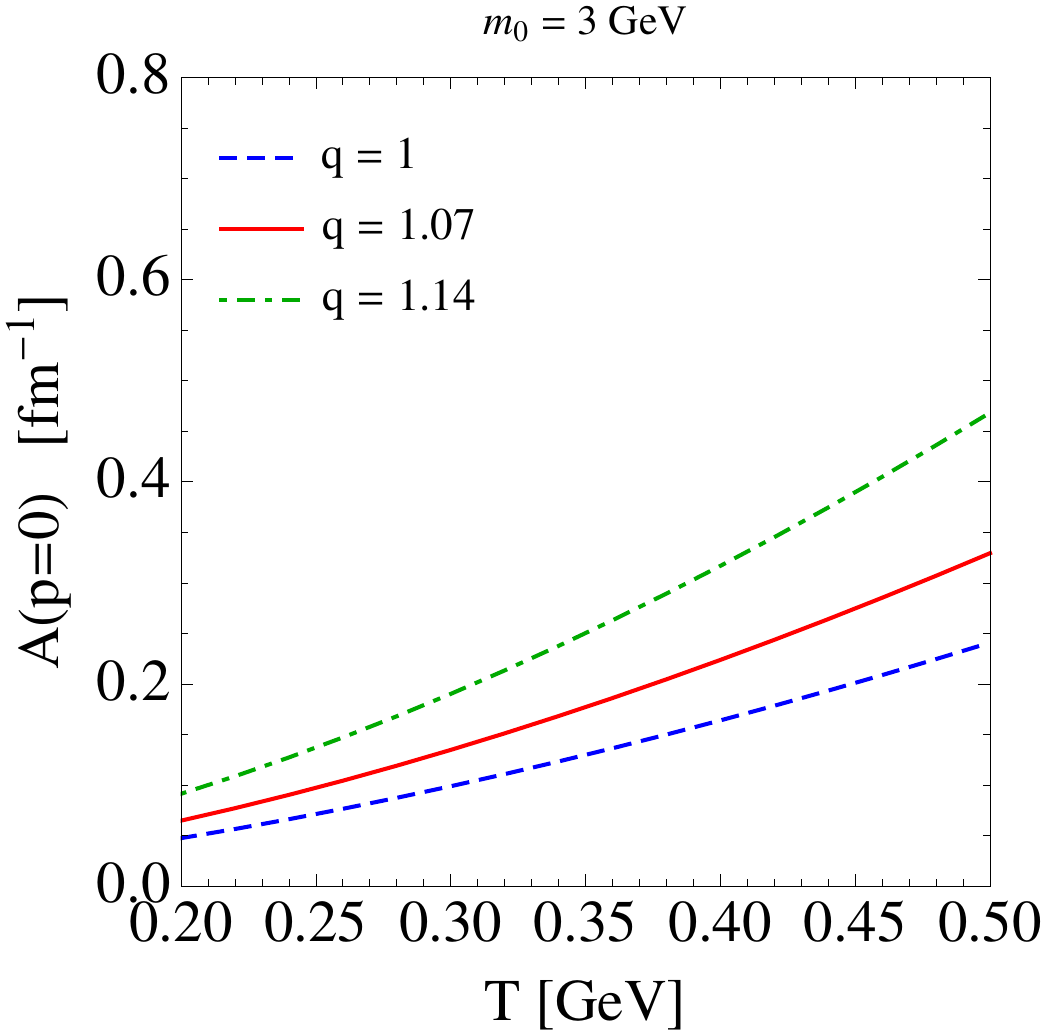}  \hspace{0.5cm} 
\includegraphics[width=0.31\textwidth]{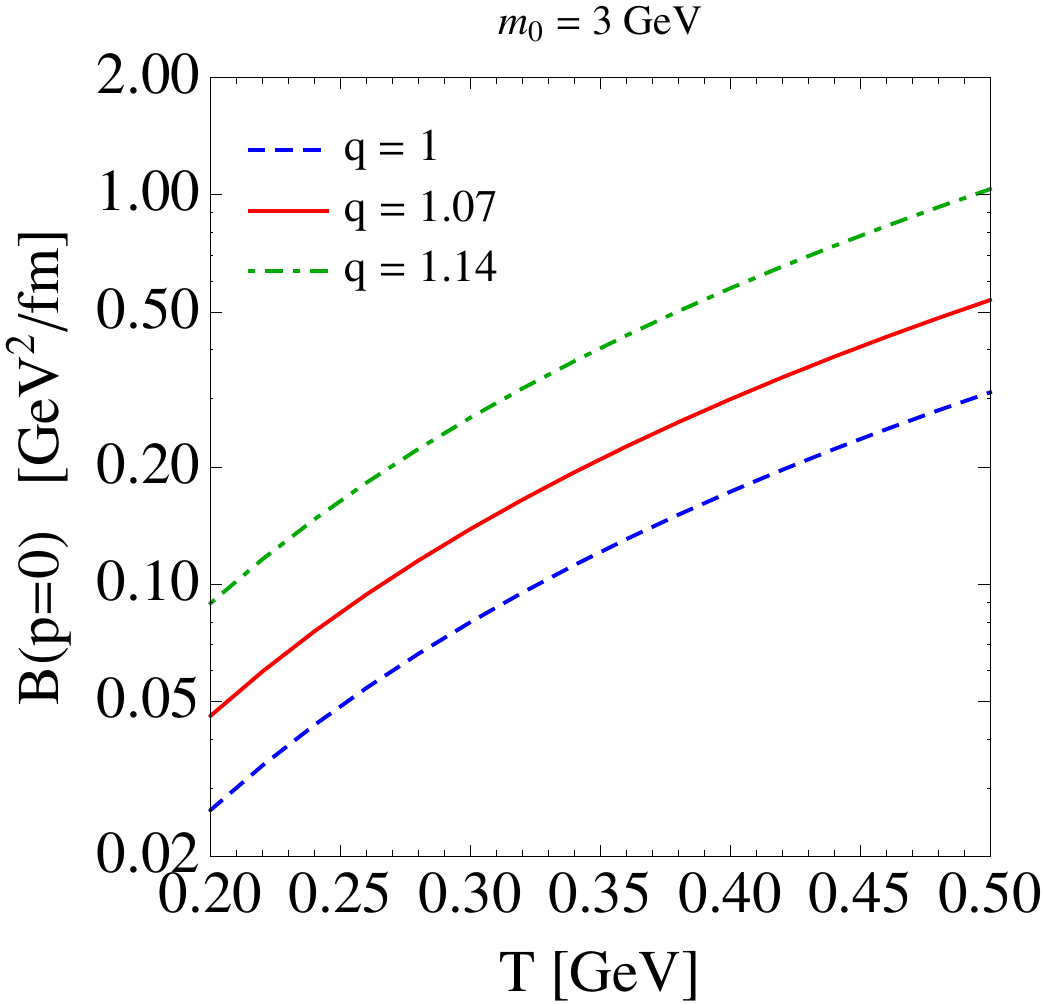}  \hspace{0.5cm} \includegraphics[width=0.30\textwidth]{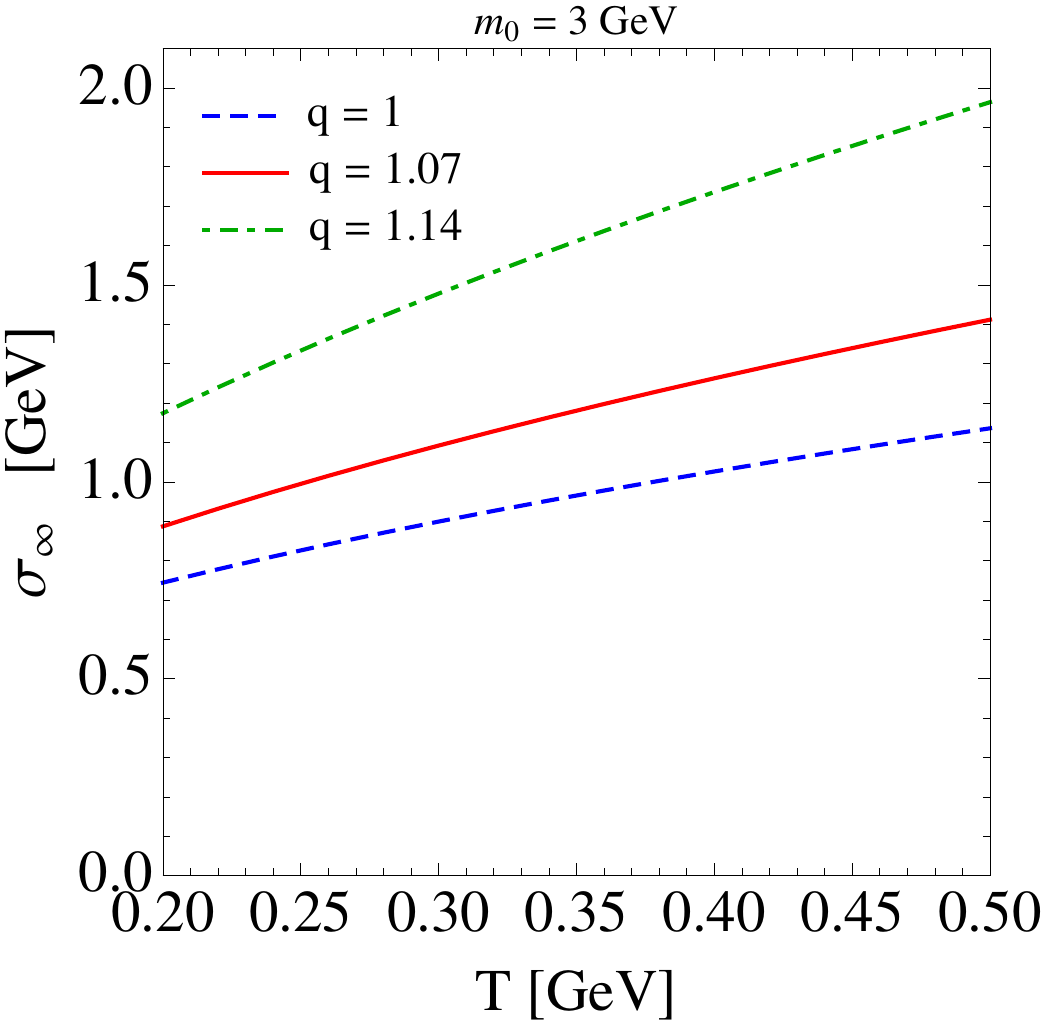} 
 \caption{Plot of the drift coefficient $A(p=0)$ (left panels), diffusion coefficient $B(p=0)$ (middle panels), and width of the distribution in equilibrium $\sigma_\infty$ (right panels) as a function of the temperature. We have displayed the results for $m_0 = 1.5 \, \textrm{GeV}$ (upper panels) and $m_0 = 3 \, \textrm{GeV}$ (lower panels), and for $q=1, 1.07$ and $1.14$. We have considered $\alpha_s = 0.6$ for the strong coupling constant, and $\mu = T$ for a regulator introduced into the internal gluon propagator in the $t$-channel-exchange diagrams to include the effects of Debye screening, cf. Ref.~\cite{Svetitsky:1987gq} for details.}
\label{fig:ABsigma}
\end{figure*}

We note that the use of the same parameters $A$ and $B$ for the PPE is inconsistent, since the distributions obtained in this case are $q$-exponentials, while we used exponentials in the calculations of the transport coefficients. The difference in the behaviour of the parameters when the non-extensive momentum distributions are used can be observed in Fig.~\ref{fig:ABsigma}, where we see that both $A(p^2)$ and $B(p^2)$ increase as $q$ increases with respect to the value $q=1$, when the Boltzmann case is recovered. However, the ratio $B/A$ increases, so the width of the distribution at the stationary regime increases with $q$. 

Comparing the results in Fig.~\ref{fig:ABsigma}, we observe that the coefficient $A$ decreases with the quark mass, $m_0$, while the coefficient $B$ increases. Therefore, the ratio $B/A$, and consequently the width at the stationary regime, increases with the quark mass.

\begin{figure*}[t]
\centering
\includegraphics[width=0.32\textwidth]{plotpM.pdf}  \hspace{1.5cm} 
\includegraphics[width=0.30\textwidth]{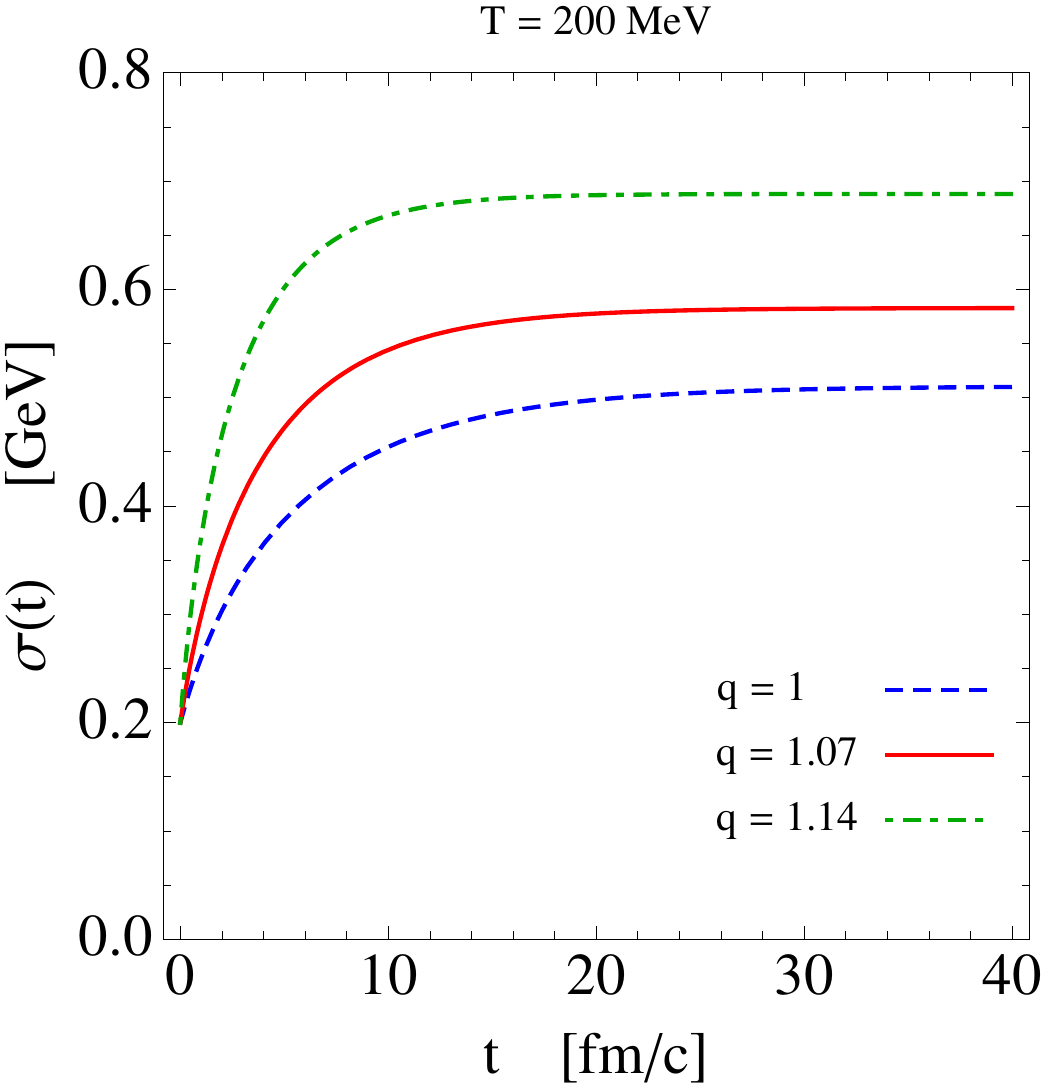} 
 \caption{Plots of $p_M(t)/p_o$ and $\sigma(t)$ as functions of time. We keep for $\sigma(t)$ the PPE solution with $q = 1.07$, while comparing the effects that produce in these parameters the $q$-dependence in $A$ and $B$ computed in Boltzmann-Gibbs statistics $(q=1)$ and in $q$-statistics $(q > 1)$. In the right panel, we have considered the initial width $\sigma_o = 0.2 \, \textrm{GeV}$.}
\label{fig:pMsigmaq}
\end{figure*}


Observe that, while the FPE solution is a Gaussian with time-dependent parameters, the solution for the PPE is a $q$-Gaussian. In both cases, the system approaches asymptotically the equilibrium, and for $t \rightarrow \infty$, the heavy-quark momentum occupies a 3D region around the null momentum region, with a distribution that is isotropic and follows a Gaussian or a $q$-Gaussian, in the cases of the FPE and of the PPE, respectively. 

Walton and Rafelski~\cite{WaltonRafelski} had compared the FPE with a non-linear Fokker-Planck Equation in order to introduce the S$_q$ and found that the results by Svetitsky for the FPE were well reproduced by a $q$-Gaussian with $q=1.1$. The results in Fig.~\ref{fig:fpz} clarify the agreement found by those authors: with that value for the entropic index the Boltzmann distribution is reproduced also by the non-extensive result.

The fact that the PPE leads to $q$-exponential distributions is in agreement with the experimental data from HEP laboratories. This result shows that PPE should be used instead of FPE. It is interesting to observe in Fig.~\ref{fig:fpz} that for temperatures around $T=200~\textrm{MeV}$ the PPE solution approaches the stationary value more slowly, but the distribution width is asymptotically similar to that of the FPE solution if $1 < q < 1.15$. For values of $q$ above that range, the width of the non-extensive distribution increases fast as the entropic index increases. For higher temperatures, though, the width of the non-extensive distributions is always larger than the distributions obtained from the FPE.

One interesting quantity is the length of the heavy-quark path in the medium, which can be calculated by using
\begin{equation}
    m \, v_M(t)=p_M(t)=p_o e^{-At}\,, \label{eqPM}
\end{equation}
where $m = m_0/ (1-v_M^2(t))^{1/2}$, and $m_0$ is the rest mass of the heavy quark. The path length until equilibrium is given by
\begin{equation}
    \ell_M=\int_0^{\infty} dt \, v_M(t) \,,
\end{equation}
resulting in
\begin{equation}
    \ell_M=\frac{1}{A} {\textrm{arcsinh}}(p_o/m_0)\,.   
    \label{eq:lM_arcsinh}
\end{equation}
Observe that for $p_o$ small compared to $m_0$, this result can be approximated to
\begin{equation}
    \ell_M=  \frac{p_o}{m_0} \frac{1}{A} \,, \label{lM} 
\end{equation}
which corresponds to the non-relativistic result $\ell_M = v_o \tau_M$, with $v_o$ being the initial velocity of the heavy quark, and $\tau_M = 1/A$ -- the relaxation time. The different evolution of the parameters of the distribution for the different values of the entropic index, $q$, are displayed in Fig.~\ref{fig:pMsigmaq}. Observe that, for any value of $q$, the behaviour of $p_M(t)$ is always exponential, as given by Eq.~(\ref{soltutionparameters}). This results from the fact that the behaviour of the parameter $p_M(t)$ in the solutions depends only on the drift term of Eqs.~(\ref{FPE}) and~(\ref{PPE}), and not on the diffusion term. Setting $B_{ij}=0$ in those equations results in the same equation for both cases, and the general solution to the new equation is 
\begin{equation}
f(\mathbf p,t)=\frac{1}{\left(\sqrt{2\pi} \sigma(t)\right)^3} ~F\left[\frac{(\mathbf p - \mathbf p_M(t))^2}{2 \sigma(t)^2} \right] \,,
\end{equation}
where $F$ is any analytical function, and the time-dependent parameters are 
\begin{equation}
 \begin{cases}
\mathbf p_{M}(t)=\mathbf p_{o} \exp[-A t] \\ \sigma(t) = \sigma_o \exp[-A t]
 \end{cases}\,.
\end{equation}
The $q$-exponential and the exponential functions are particular solutions of the simplified equation.

Hence, for the same initial momentum, the length will be proportional to $A^{-1}$. Since, as shown above, the non-additive statistics leads to larger values of $A$ as $q$ increases, the minimal size of the medium for allowing the heavy quark to reach the stationary state diminishes with $q$. The experimental data from high energy collisions indicate that the heavy quark reaches the stationary state inside the medium before the freeze-out. These indications are the momentum distributions of the heavy hadrons and the collective flow observed through heavy hadrons, which are similar to those observed for pions.

The results presented in the previous section indicate that the best situation to compare the results associated with heavy quark time or distance travelled to reach the stationary state, corresponds to relatively low-mass quarks as this is  better to distinguish between FPE or PPE. This is because $A$ decreases with the quark mass, so the length $\ell_M$ is larger in this case. 
\begin{figure}[t]
\centering
\includegraphics[width=0.45\textwidth]{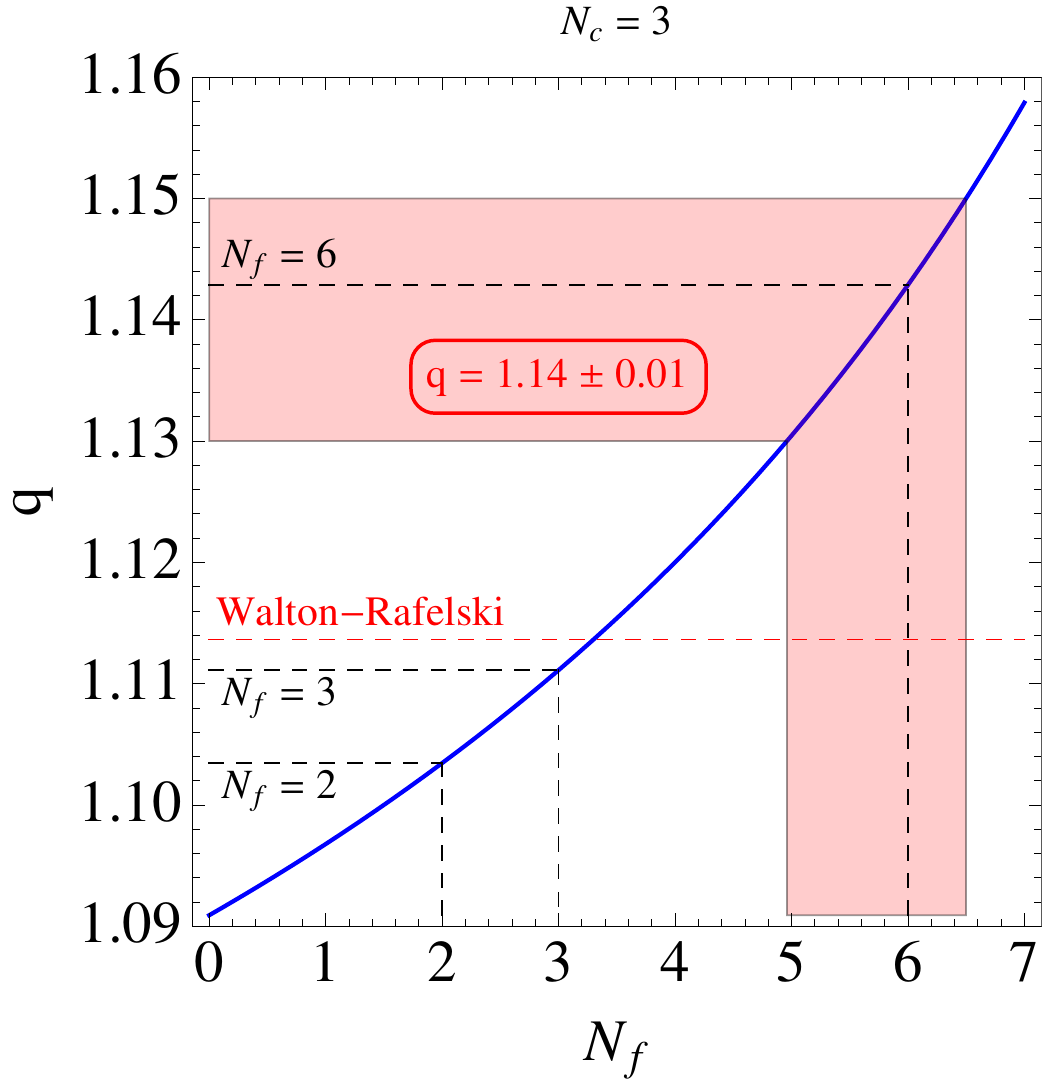} 
 \caption{Plot of the entropic index $q$ as a function of $N_f$ (solid blue line), as given by Eq.~(\ref{ndof})  with $N_c = 3$; for $N_f=2, 3, 6$ we respectively obtain $q=32/29\simeq1.10$, $q=10/9\simeq1.11$ and $q=8/7\simeq 1.14$. The experimental LHC/CERN value~\cite{Wong:2015mba} is $q=1.14 \pm 0.01$, while the Walton-Rafelski value \cite{WaltonRafelski} is $q=1+1/8.8 \simeq 1.114$ (red dashed line). We display as a shaded (red) area the region corresponding to $q = 1.14 \pm 0.01$ and, as horizontal lines, the values of $q$ for $N_f = 2, 3$ and $6$.}
\label{fig:qNf}
\end{figure}

The investigation of those physical quantities associated with the width of the distribution can be suitably studied by observing the properties of heavier quarks. In particular, the difference in the stationary distribution width as a function of the entropic index $q$ is enhanced for heavier quarks.
The results obtained above open an opportunity to test the number of degrees of freedom ($n_{dof}$) relevant to the interaction of the heavy quark with the medium. The number of colours, $N_c$, and the number of flavours, $N_f$, are related to the number of degrees of freedom by~\cite{Deppman:2019yno}
\begin{equation}
  n_{dof}=\frac{1}{q-1}=\frac{11}{3}N_c-\frac{4}{3}\frac{N_f}{2} \,. \label{ndof}
\end{equation}
Considering $N_c=3$ and $N_f=6$, we get $n_{dof}=7$ and $q=8/7 \simeq 1.14$. However, the theory that leads to Eq.~(\ref{ndof}) assumes that the masses of the quarks are negligible. In the case where the quark masses become relevant, due to the $q$-exponential behaviour of the distributions, their contributions to the thermalization of the quark can be neglected, which corresponds to a reduction of the number of effective flavours in the process. If we use $N_f=2$ in Eq.~(\ref{ndof}), we get $n_{dof}=29/3 \sim 9.7$ and $q=32/29 \simeq 1.10$. If one considers $N_f = 3$, then $q=10/9\simeq 1.11$. We display in Fig.~\ref{fig:qNf} the behaviour of the entropic index $q$ with $N_f$ as given by Eq.~(\ref{ndof}).

Walton and Rafelski~\cite{WaltonRafelski} used a non-linear FPE associated with the non-additive statistics to compare with Svetitsky's calculations~\cite{Svetitsky:1987gq}, and found $q=1.1$, which is in agreement with the scenarios for $N_f=3$ and for $N_f=2$. In Fig.~\ref{fig:qNf} we plot the value for $q$ as a function of $N_f$ for $N_c = 3$. The shaded area corresponds to the interval for $q$ that results from experimental data analysis by HEP experiments where the QGP deconfined quark regime is attained. Calculations that consider only two or three flavours, as those carried out here, lead to different values for $q$, in accordance with the predicted in Eq.~(\ref{ndof}). 

The meaning implied by Eq.~(\ref{ndof}) can be clearly understood in face of the present results. That formula was obtained in Ref.~\cite{Deppman:2019yno}, where the quark masses are considered negligible. For processes at relatively low energy, where the quark masses become relevant, the finite mass effects must be taken into account, and one consequence is the dump of the heavy quark contributions due to the $q$-exponential behaviour of the particle spectra in the QGP. A phenomenological method for taking this effect into account is by reducing the effective number of flavours for the process, as commonly used in QCD calculations. The results obtained in the present work and that obtained in Ref.~\cite{WaltonRafelski} confirm the predictions given by that equation.

One of the main differences in the solutions of the FPE with respect to the solutions of the PPE is the width of the distribution, which becomes larger for the PPE than for the FPE as soon as the interaction between the heavy-quark and the medium starts, and increases continuously until the stationary regime is reached.  

The heavy quark dynamic has implications in many aspects of the high-energy collisions~\cite{PasechnikSumbera,Adolfsson:2020dhm}. The jet quenching~\cite{Qin:2015srf,Apolinario:2015bfm} and the heavy hadron suppression~\cite{Casalderrey-Solana:2018wrw} are phenomena that can benefit from the results obtained here. Due to the different relaxation times and displacement in the medium, the different behaviours of the additive and non-additive solutions can be important for determining the dynamical effects in small and large systems.

\begin{table}[H]
\centering
\caption{Table of physical quantities for $T=200$ MeV and $m_0 = 1.5$ GeV. We have assumed $v_o = 0.3 c$ as the initial velocity of the heavy quark in the computation of $\ell_M$. The quantities $\sigma_{\infty}$ and $\ell_M$ are calculated according to Eqs.~(\ref{sigmainft}) and~(\ref{eq:lM_arcsinh}), respectively, and $\tau_M=1/A$. Here, $A$ and $B$ are defined in Eqs.~(\ref{eqA}) and (\ref{eqB}), and considered to be approximately constant. \\} \label{Table}
\begin{tabular}{| c | c | c |}
\cline{2-3}
 \multicolumn{1}{c|}{} & \textbf{FPE } & \textbf{PPE } \\ 
  \multicolumn{1}{c|}{} & \textbf{Eq. (1)} & \textbf{Eq. (2) with q=1.10} {\footnotesize 
 } \\ \hline 
 & $A = 0.0820$ fm$^{-1}$ & $A = 0.0820$ fm$^{-1}$ \\
{\bf Eqs. (\ref{eqA}), (\ref{eqB})} & $B = 0.0216$ GeV$^{2}$/\textrm{fm} & $B = 0.0216$ GeV$^{2}$/\textrm{fm}  \\
 & $\sigma_{\infty} = 0.513$ GeV & $\sigma_{\infty} = 0.510$ GeV\\
 {\bf with q=1} & $\tau_M$ = 12.20 fm/c & $\tau_M = 12.20$ fm/c \\
   & $\ell_M = 3.78$ fm & $\ell_M = 3.78$ fm    \\
  \hline
  & $A = 0.129$ fm$^{-1}$ & $A = 0.129$ fm$^{-1}$ \\
{\bf Eqs. (\ref{eqA}), (\ref{eqB})}  & $B = 0.0486$ GeV$^{2}$/\textrm{fm}  & $B = 0.0486$ GeV$^{2}$/\textrm{fm} \\
  & $\sigma_{\infty}$ = 0.613 GeV   & $\sigma_{\infty}$ = 0.629 GeV \\
  {\bf with q=1.10}  &  $\tau_M$ = 7.74 fm/c &  $\tau_M = 7.74$ fm/c \\
  &  $\ell_M = 2.39$ fm &  $\ell_M = 2.39$ fm \\  
\hline
\end{tabular}
\end{table}

Table~\ref{Table} summarizes the results obtained in the analysis performed here, displaying the calculations of the physical parameters $\sigma_{\infty}$, $\tau_M$ and $\ell_M$ corresponding, respectively, to the stationary distribution width, the relaxation time and the characteristic distance travelled by the heavy quark till the equilibrium. The results for FPE and for PPE are shown. The top-left and the bottom-right panels show fully consistent calculations: the first corresponds to additive coefficients used in the additive FPE. The second corresponds to non-additive coefficients and non-additive PPE.


In this work, we have studied the dynamical properties of heavy quarks in the medium as predicted by the Fokker-Planck Equation and by its non-additive counterpart, the Plastino-Plastino Equation. The transport coefficients were obtained for both the additive and the non-additive statistics through a microscopic second-order calculation of the heavy quark interaction with the light quarks and gluons in the medium.

We observed different dynamical characteristics in the results of the additive and non-additive equations. The main effect of introducing the Plastino-Plastino equation is a broader distribution in the stationary regime, which is obtained even when the transport coefficients calculations do not include the non-additive features.

When the non-additive effects are included in the transport coefficients, by considering the non-extensive distributions, the differences in the dynamics of the quarks are more evident. Not only the width of the distribution is broader, but also the relaxation time and the quark displacement until the stationary regime is reached are smaller than the equivalent quantities in the additive case. We emphasize that the only  fully consistent calculations are those with FPE and transport coefficients calculated with $q=1$, and the PPE equations using the transport coefficients calculated with the same value for $q \ne 1$. These cases correspond to the top-left and bottom-right cases in Table~\ref{Table}. The comparison between the results obtained here and that performed in Ref.~\cite{WaltonRafelski} corroborates the findings of Ref.~\cite{Deppman:2019yno}, where the fractal approach to the non-perturbative QCD led to Eq.~(\ref{ndof}) relating the entropic index $q$ to the number of colours, $N_c$ and the number of flavours, $N_f$.

The results obtained here have implications in the study of several aspects of high-energy collisions. Phenomena like jet quenching and heavy hadron suppression depend on a careful description of the quark dynamics and can benefit from these results. A possible way to  distinguish experimentally between the FPE and the PPE solutions is by observing the distance travelled by a heavy quark in the medium. This can be done experimentally by observing the ratio between di-jets carrying heavy quarks and non-correlated jets as a function of collision centrality. For less central collisions, the relative number of di-jets should increase, and the size of the system when this happens should be related to the distance travelled by the heavy quark in the medium. The study of the momentum distribution of jets is an alternative way to investigate the phenomenon, and this will be calculated in future work. 

Although the work performed here was motivated by the study of QGP and HEP, the results can be of interest to researchers in many areas, such as solar plasma~\cite{CORADDU2003473}, ionic diffusion~\cite{Curilef} and neutron star~\cite{Annala:2019puf,Annala:2020rgx,Cardoso2017,Sen:2021tdu}, for instance. The micro-calculation of the transport coefficients for hadronic matter has been under debate~\cite{Sen:2021tdu}, with implications in hydrodynamical models. The results obtained here imply modifications in both the form of the solution for the dynamical equation and on the transport coefficients, and can be of interest to researchers in those fields.

\section{Acknowledgements}

E.M. and A.D. thank E. Ruiz Arriola for discussions. A.D. would like to thank the University of Granada, where part of this work has been done, for the hospitality and financial support under a grant of the Visiting Scholars
Program of the Plan Propio de Investigaci\'on of the University of Granada. He also acknowledges the hospitality at Carmen de la Victoria. The work of E.M. is supported by the project
PID2020-114767GB-I00 funded by MCIN/AEI/10.13039/501100011033, by
the FEDER/Junta de Andaluc\'{\i}a-Consejer\'{\i}a de Econom\'{\i}a y
Conocimiento 2014-2020 Operational Program under Grant A-FQM178-UGR18, and by Junta de Andaluc\'{\i}a under Grant FQM-225. The research of E.M. is also supported by the Ram\'on y Cajal Program of the Spanish MCIN under Grant RYC-2016-20678. A.D. is supported by the Project INCT-FNA (Instituto Nacional de Ci\^encia e Tecnologia - F\'{\i}sica Nuclear Aplicada) Proc. No. 464898/2014-5, by the Conselho Nacional de Desenvolvimento Cient\'{\i}fico e Tecnol\'ogico (CNPq-Brazil), grant 304244/2018-0, by
Project INCT- FNA Proc. No. 464 898/2014-5, and by FAPESP, Brazil
grant 2016/17612-7. R.P. is supported in part by the Swedish Research Council grants, contract numbers 621-2013-4287 and 2016-05996, as well as by the European Research Council (ERC) under the European Union’s Horizon 2020 research and innovation programme (grant agreement No. 668679). C.T. is partially supported by CNPq and Faperj (Brazilian agencies).


\begin{thebibliography}{10}

\bibitem{Hees-Rapp}
H.~van Hees and R.~Rapp, ``{Thermalization of heavy quarks in the quark-gluon
  plasma},'' {\em Phys. Rev. C}, vol.~71, p.~034907, 2005.

\bibitem{He:2022ywp}
M.~He, H.~van Hees, and R.~Rapp, ``{Heavy-Quark Diffusion in the Quark-Gluon
  Plasma},'' arXiv:2204.09299 2022.

\bibitem{Das-Alam-Mohanty}
S.~K. Das, J.-e. Alam, and P.~Mohanty, ``{Probing quark gluon plasma properties
  by heavy flavours},'' {\em Phys. Rev. C}, vol.~80, p.~054916, 2009.

\bibitem{Svetitsky:1987gq}
B.~Svetitsky, ``{Diffusion of charmed quarks in the quark-gluon plasma},'' {\em
  Phys. Rev. D}, vol.~37, pp.~2484--2491, 1988.

\bibitem{Tsallis}
C.~Tsallis, ``{Possible Generalization of the Boltzmann-Gibbs Statistics},''
  {\em {Journal of Statistical Physics}}, vol.~{52}, no.~{1-2}, pp.~{479--487},
  {1988}.

\bibitem{Marques-Cleymans-Deppman-2015}
L.~Marques, J.~Cleymans, and A.~Deppman, ``{Description of high-energy pp
  collisions using Tsallis thermodynamics: Transverse momentum and rapidity
  distributions},'' {\em {Physical Review D}}, vol.~{91}, no.~{5}, {2015}.

\bibitem{Marques-Andrade-Deppman-2013}
L.~Marques, E.~Andrade, II, and A.~Deppman, ``{Nonextensivity of hadronic
  systems},'' {\em {Physical Review D}}, vol.~{87}, no.~{11}, {2013}.

\bibitem{WilkWlodarkzyk-multiparticle}
G.~Wilk and Z.~Wlodarczyk, ``Some intriguing aspects of multiparticle
  production processes,'' {\em International Journal of Modern Physics A},
  vol.~33, APR 10 2018.

\bibitem{TsallisBook}
C.~Tsallis, {\em Introduction to Nonextensive Statistical Mechanics -
  Approaching a Complex World}.
\newblock Springer, 2009 - Second Edition 2023.

\bibitem{PLASTINO1995347}
A.~Plastino and A.~Plastino, ``Non-extensive statistical mechanics and
  generalized fokker-planck equation,'' {\em Physica A: Statistical Mechanics
  and its Applications}, vol.~222, no.~1, pp.~347--354, 1995.

\bibitem{Muskat}
M.~Muskat, {\em The Flow of Homogeneous Fluids Through Porous Media}.
\newblock McGraw-Hill, 1937.

\bibitem{Schwammle}
V.~Schwämmle, F.~Nobre, and C.~Tsallis, ``q-gaussians in the porous-medium
  equation: stability and time evolution.,'' {\em The European Physical Journal
  B}, vol.~66, p.~537–546, 2008.

\bibitem{Schwammle2009}
V.~Schwämmle, E.~M.~F. Curado, and F.~D. Nobre, ``Dynamics of normal and
  anomalous diffusion in nonlinear fokker-planck equations,'' {\em The European
  Physical Journal B}, vol.~70, pp.~107--116, Jul 2009.

\bibitem{WaltonRafelski}
D.~B. Walton and J.~Rafelski, ``{Equilibrium distribution of heavy quarks in
  Fokker-Planck dynamics},'' {\em Phys. Rev. Lett.}, vol.~84, pp.~31--34, 2000.

\bibitem{Wong:2015mba}
C.-Y. Wong, G.~Wilk, L.~J.~L. Cirto, and C.~Tsallis, ``{From QCD-based
  hard-scattering to nonextensive statistical mechanical descriptions of
  transverse momentum spectra in high-energy $pp$ and $p\bar p$ collisions},''
  {\em Phys. Rev. D}, vol.~91, no.~11, p.~114027, 2015.

\bibitem{Deppman:2019yno}
A.~Deppman, E.~Megias, and D.~P. Menezes, ``{Fractals, nonextensive statistics,
  and QCD},'' {\em Phys. Rev. D}, vol.~101, no.~3, p.~034019, 2020.

\bibitem{PasechnikSumbera}
R.~Pasechnik and M.~Šumbera, ``Phenomenological review on quark–gluon
  plasma: Concepts vs. observations,'' {\em Universe}, vol.~3, no.~1, 2017.

\bibitem{Adolfsson:2020dhm}
J.~Adolfsson {\em et~al.}, ``{QCD challenges from pp to A\textendash{}A
  collisions},'' {\em Eur. Phys. J. A}, vol.~56, no.~11, p.~288, 2020.

\bibitem{Qin:2015srf}
G.-Y. Qin and X.-N. Wang, ``{Jet quenching in high-energy heavy-ion
  collisions},'' {\em Int. J. Mod. Phys. E}, vol.~24, no.~11, p.~1530014, 2015.

\bibitem{Apolinario:2015bfm}
L.~Apolin\'ario, N.~Armesto, G.~Milhano, and C.~A. Salgado, ``{In-medium jet
  evolution: interplay between broadening and decoherence effects},'' {\em
  Nucl. Phys. A}, vol.~956, pp.~681--684, 2016.

\bibitem{Casalderrey-Solana:2018wrw}
J.~Casalderrey-Solana, Z.~Hulcher, G.~Milhano, D.~Pablos, and K.~Rajagopal,
  ``{Simultaneous description of hadron and jet suppression in heavy-ion
  collisions},'' {\em Phys. Rev. C}, vol.~99, no.~5, p.~051901, 2019.

\bibitem{CORADDU2003473}
M.~Coraddu, M.~Lissia, G.~Mezzorani, and P.~Quarati, ``Super-kamiokande hep
  neutrino best fit: a possible signal of non-maxwellian solar plasma,'' {\em
  Physica A: Statistical Mechanics and its Applications}, vol.~326, no.~3,
  pp.~473--481, 2003.

\bibitem{Curilef}
S.~Curilef, ``Derivation and analytical solutions of a non-linear diffusion
  equation applied to non-constant heat conductivity and ionic diffusion in
  glasses,'' {\em Chaos: An Interdisciplinary Journal of Nonlinear Science},
  vol.~32, no.~11, p.~113133, 2022.

\bibitem{Annala:2019puf}
E.~Annala, T.~Gorda, A.~Kurkela, J.~N\"attil\"a, and A.~Vuorinen, ``{Evidence
  for quark-matter cores in massive neutron stars},'' {\em Nature Phys.},
  vol.~16, no.~9, pp.~907--910, 2020.

\bibitem{Annala:2020rgx}
E.~Annala, {\em From QCD to Neutron Stars and Back : Probing the Fundamental
  Properties of Dense Matter}.
\newblock PhD thesis, Helsinki U., 2020.

\bibitem{Cardoso2017}
P.~H.~G. Cardoso, T.~Nunes~da Silva, A.~Deppman, and D.~P. Menezes, ``Quark
  matter revisited with non-extensive mit bag model,'' {\em The European
  Physical Journal A}, vol.~53, no.~10, p.~191, 2017.

\bibitem{Sen:2021tdu}
D.~Sen, N.~Alam, and S.~Ghosh, ``Estimation of transport coefficients of dense
  hadronic and quark matter,'' {\em Chinese Physics C}, 2023 - accepted for
  publication. doi: 10.1088/1674-1137/acb992.

\end{thebibliography}

\end{document}